\documentclass[manuscript=article, journal=jpcafh]{achemso}

\usepackage[version=4]{mhchem}
\usepackage{ulem}
\usepackage{siunitx}
\DeclareSIUnit\calo{cal}
\DeclareSIUnit\Debye{D}
\DeclareSIUnit\Molar{\textsc{m}}
\DeclareSIUnit\hartee{H} 
\DeclareSIUnit\angstrom{\text {Å}}

\usepackage{subcaption}
\usepackage{tcolorbox}

\title{Strong-Coupling Modification of Singlet-Fission Dynamical Pathways}

\author{Lisamaria Wallner}
\email{l.wallner@stud.uni-heidelberg.de}
\affiliation{Theoretische Chemie,
             Physikalisch-Chemisches Institut,
             Universität Heidelberg,
             Im Neuenheimer Feld 229, 69120 Heidelberg, Germany}
\author{Charlotte Remnant}
\affiliation{Theoretische Chemie,
             Physikalisch-Chemisches Institut,
             Universität Heidelberg,
             Im Neuenheimer Feld 229, 69120 Heidelberg, Germany}
\author{Oriol Vendrell}
\affiliation{Theoretische Chemie,
             Physikalisch-Chemisches Institut,
             Universität Heidelberg,
             Im Neuenheimer Feld 229, 69120 Heidelberg, Germany}
\alsoaffiliation{Interdisciplinary Center for Scientific Computing,
                 Universität Heidelberg,
                 Im Neuneheimer Feld 205, 69120 Heidelberg, Germany}
\email{oriol.vendrell@uni-heidelberg.de}

\date{\today}

\begin{document}
\maketitle
\begin{abstract}
We investigate theoretically the influence of strong light-matter coupling on
    the initial steps of the photo-triggered singlet-fission process.  In
    particular we focus on intra-molecular singlet fission in a TIPS-pentacene
    dimer derivative described by a vibronic Hamiltonian including the optically
    active singlet excited states, doubly excited and charge transfer states, as
    well as the final triplet-triplet pair state.  Quantum dynamics simulations
    of up to four dimers in the cavity indicate that the modified resonance
    condition imposed by the cavity strongly quenches the passage through the
    intermediate charge transfer and double-excitation states, thus largely
    reducing the triplet-triplet yield in the bare system.  Subsequently, we
    modify the system parameters and construct a model Hamiltonian where the
    optically-active singlet excitation lies below the final triplet-triplet
    state such that the yield of the bare system becomes insignificant.  In this
    case we find that using the upper polariton as the doorway state for
    photo-excitation can lead to a much enhanced yield. This pathway is
    operative provided that the system is sufficiently rigid to prevent vibronic
losses from the upper polariton to the dark-states manifold.
\end{abstract}

\section{Introduction}

Recent efforts in the understanding of strong light-matter interaction in
optical cavities have brought to light diverse field-induced modifications
of chemical properties, including modified thermal chemistry, photochemistry, as
well as transport phenomena in materials.\cite{byrnes_excitonpolariton_2014,
coles_polariton-mediated_2014, mewes_energy_2020,
zhong_non-radiative_2016, fidler_ultrafast_2023, satapathy_selective_2021}
These modifications are attributed to the formation of light-matter hybrid
states, so called polaritons.
Exciton-polaritons, thus formed by the combination of electronic excitations
with light, have been shown to alter the properties of the bare excitonic
systems, influencing processes like electron
transport\cite{balasubrahmaniyam_enhanced_2023, orgiu_conductivity_2015,
sokolovskii_multi-scale_2023},
light-harvesting\cite{esteso_light-harvesting_2021, polak_manipulating_2020} or
energy transfer \cite{zhong_non-radiative_2016,coles_polariton-mediated_2014}.
Among these, singlet fission (SF) is an important
photophysical process for light-harvesting applications that can be used to
circumvent the Shockley-Queisser efficiency limit thereby increasing the
external quantum efficiency of organic solar cells.\cite{shockley_detailed_2004,
smith2010singlet,congreve2013external}
In SF a singlet exciton (SE) converts into a multiexcitonic (ME) state
consisting of a correlated triplet pair state with singlet character, thus being
a spin-allowed process where least two molecules, or molecular fragments in the
case of internal singlet fission (iSF), must be involved.
The triplet pair state eventually decoheres and separates to form two
independent triplet excitons localized on two different molecules or
molecular
fragments.\cite{pensack_observation_2016,burdett_dynamics_2013,
zirzlmeier2015singlet}
A wide exploitation of SF for solar energy conversion remains
challenging, though, as only a
limited number of materials can display this
phenomenon.\cite{casanova2018theoretical, padula_singlet_2019, smith2013recent}
In particular, monomers with good characteristics for SF must feature a
triplet state at an energy slightly below half of the first singlet excitation,
such that one singlet excited state can transfer into two separate triplets
resonantly.
Unfortunately, it is challenging to control this energetic boundary condition
intrisically, since it depends to a large extent on the electronic
structure of the corresponding monomers.

Therefore, the utilization of exciton-polaritons represents an appealing
strategy to modify the SF characteristics of materials.
A number of experimental studies have already been
conducted on how strong light-matter coupling affects the SF process in
different types of cavities.\cite{liu_role_2020, takahashi2019singlet,
polak_manipulating_2020, ptheurer_strong_2023,
berghuis_enhanced_2019, kolesnichenko_charge-transfer_2024}.
For instance, amorphous rubrene thin films under
strong-coupling showcased a field-induced decoupling effect, thereby suppressing
SF.\cite{takahashi2019singlet}
Additionally, polariton-mediated enhancement of
light harvesting in TIPS-tetracene thin films was observed.
\cite{polak_manipulating_2020}
Furthermore, studies have been conducted that examine the interaction
between polaritonic effects and SF near metallic surfaces.
The findings indicate that the dynamics remained unchanged in the
presence of
polaritons.\cite{kolesnichenko_charge-transfer_2024, kolesnichenko_hot_2024}

Theoretical studies have investigated the impact of strong coupling on SF
systems, with a particular focus on the direct SF pathways within fundamental
three-state models, which comprise the ground state (GS), singlet excited state
(SE), and the final multiexcitonic (ME)
state.\cite{martinez-martinez_polariton-assisted_2018, sun_engineering_2022,
gu_optical-cavity_2021}
Higher-lying
electronic states, particularly the charge transfer (CT) states,
have been included in some of the polaritonic models
in recognition
of their significance in the SF process. \cite{zhang_joint_2021,
climent_not_2022}
These states, in addition to the high-lying doubly excited states, play a
crucial role in the SF process through indirect pathways.
\cite{reddy_intramolecular_2018}

In order to engineer the optimal passage through these dynamical pathways,
the effect of the cavity on the transient population of the intermediate states
needs to be well understood. In addition, the coupling to the cavity results in
new bright resonances centered around the lower and the upper polaritonic
states, which can be used as doorway states to initiate the SF process.
The
specificities of starting through either resonance can be investigated by
considering explicit laser pulses tuned to specific resonances of
the absorption spectrum~\cite{ulusoy_modifying_2019}.

Hence, here we consider a realistic model for iSF
that incorporates both the high-lying charge transfer (CT) and doubly
excited (DE) states, with the objective of comprehensively capturing their
influence on the modified SF mechanism through full quantum dynamics simulations and the
consideration of collective effects.
Through the inclusion of the femtosecond excitation laser-pulse tuned to
specific resonances in the model we can carefully investigate the how the
different doorway states result in different dynamics and finally different yields.
Additionally, we examine in some detail the possibility of using
the upper polariton as a doorway. This strategy may be beneficial for
materials where the first singlet excited state is energetically too low
compared to twice the triplet excitation energy, which is the case for smaller
organic systems. A significant number of molecular
systems exhibit this energy distribution,\cite{padula_singlet_2019}
and in general smaller molecules may be more robust towards degradation
processes and easier to prepare and produce.

\section{Theory and Methods}

\subsection{Polaritonic-Molecular Hamiltonian}

We consider $N$ molecules strongly coupled to a cavity mode and described by
the Hamiltonian
\begin{align}
    \label{eq:hamtot}
 \hat{H} = \sum^N_{n=1}\hat{H}^{(n)}_\text{mol}
                     + \hat{H}_\text{cav}
                     + \hat{H}_\text{las},
\end{align}
where $\hat{H}^{(n)}_\text{mol}$ is the $n$-th molecular
Hamiltonian,
$\hat{H}_\text{cav}$ is the cavity Hamiltonian and
$\hat{H}_\text{las}$ describes a laser pulse interacting with the system.
The molecular Hamiltonian is considered in its general diabatic representation
\begin{align}
    \label{eq:hmol}
       \hat{H}_{mol}^{(n)} &= \hat{T}^{(n)}
       + \sum_{i,j=1}^{N_e} |i_n\rangle W_{ij}(\mathbf{Q}^{(n)}) \langle j_n|,
\end{align}
where $N_e$ is the number of intramolecular electronic states considered for
each molecule,
\begin{align}
    \label{eq:diab}
    W_{ij}(\mathbf{Q}^{(n)}) =
    \langle \varphi_i^{(n)}|
        \hat{H}_\text{el}(\mathbf{r}^{(n)},\mathbf{Q}^{(n)})
    |\varphi_j^{(n)}\rangle
\end{align}
are the matrix elements of the electronic Hamiltonian in the basis of
local diabatic states for the $n$-th molecule, and the brackets indicate
integration over the local electronic coordinates $\mathbf{r}^{(n)}$.
The coupling to the quantized electromagnetic mode of the cavity is considered
in Coulomb gauge and length form
\cite{flick_atoms_2017, vendrell_collective_2018}
\begin{align}
    \label{eq:hcav}
        \hat{H}_\text{cav}
        &= \hbar \omega_c \left(\hat{a}^{\dagger} \hat{a} \right)
        +
        \sqrt{\frac{\hbar\omega_c}{2}} \lambda
        \vec{\varepsilon_c} \vec{\hat{D}}
        \left(\hat{a}^{\dagger} + \hat{a} \right)
        + \frac{1}{2}\left(\lambda \vec{\varepsilon_c} \vec{\hat{D}}\right)^{2},
\end{align}
where $\omega_c$ is the cavity frequency.
The second term describes the linear interaction between the molecule
and the cavity, where $\vec{\varepsilon_c}$ is the polarization
vector of the cavity mode and $\lambda=\sqrt{1/\varepsilon_0 V}$
is the coupling strength with quantization volume $V$
and vacuum polarization $\varepsilon_0$.
As usual, to facilitate comparisons we introduce the coupling parameter
$g=\lambda\sqrt{\hbar\omega_c/2}$ with units of electric field.
$\vec{\hat{D}}=\sum_{n=1}^{N}\vec{\hat{\mu}}$ is the total dipole operator of
the molecular ensemble.

In case we need to describe the interaction with an external laser pulse $k$
explicitly, we introduce it for convenience through its vector potential,
$\vec{E}_k(t)=-\partial \vec{A}_k(t)/\partial t$,
\begin{align}
    A_k(t)= \vec{\epsilon}_k \frac{E_{k,0}}{\omega_k}
        \exp{\left[-\frac{2\mathrm{ln}2}{F^2}\left(t-\tau_{k}\right)^2\right]
        \cos{\left(\omega_k(t-\tau_{k})\right)}},
\end{align}
where $E_{k,0}$ is the field amplitude, $\omega_k$ and $\vec{\epsilon}_k$ are
the carrier frequency and polarization direction of the laser pulse,
$F$ corresponds to the FWHM of the intensity profile, and
$\tau_{k}$ is the center of the pulse. The corresponding interaction Hamiltonian in the
dipole approximation then reads
$\hat{H}_\text{las} = - \vec{\hat{D}} \cdot \vec{E}_k(t)$.

Hamiltonian~\ref{eq:hcav} corresponds to the Pauli-Fierz
description of light-matter interaction in the case that the molecules
interact with a single effective electromagnetic mode.
From the perspective of the material system, the main approximation in
our treatment corresponds to
the introduction of a local basis
of field-free molecular states for each monomer, on which all relevant operators
are represented.
For a given set of electromagnetic modes and
coupling parameter $\lambda$, the scheme outlined above
is formally exact as long as the local molecular basis can
be considered to be complete.

\subsection{Intramolecular Singlet Fission Model in a Cavity}

The simulations on iSF reported here are based on the vibronic Hamiltonian of
\textit{o}-bis(13-(methylethynyl)pentacen-6-yl)ethynyl)benzene dimer
(\textit{o}-TIPSPm) parametrized by Reddy and
Thoss~\cite{reddy_intramolecular_2018}, which they simulated using the
multilayer multiconfiguration time-dependent Hartree method
(ML-MCTDH)~\cite{wan_03_1289,man_08_164116,vendrell_multilayer_2011}.
This model considers the 8 most important electronic states in the iSF process
of the \textit{o}-TIPSm, namely the electronic ground state (GS), the first two
locally excited states (LE), two high-lying charge transfer states (CT) and two
doubly excited states (DE), as well as the final multiexcitonic (TT) state.
The diabatic electronic Hamiltonian for this window of electronic states at the
Franck-Condon geometry $\mathbf{Q}^{(n)}=0$ is reproduced from
Ref.~\citenum{reddy_intramolecular_2018} in Table~\ref{tab:elHam}.
Their model also considers the 8 fundamental vibrational modes for the iSF
process and their respective linear and quadratic vibronic couplings, selected
by Reddy and Thoss~\cite{reddy_intramolecular_2018}, to account for the
non-adiabatic effects. The vibrations consist of mostly ring-breathing and
-stretching modes that involve the pentacene subunits. The LE states of this
model are then coupled to a resonant cavity photon according to
Eq.~\ref{eq:hcav}.
\begin{table}[t]
    \centering
\caption{Diabatic electronic Hamiltonian matrix of \textit{o}-TIPSPm
    at the FC point, $\mathbf{Q}^{(n)}=0$. Energies in $[\mathrm{meV}]$
    reproduced from Ref. \citenum{reddy_intramolecular_2018}}
\label{tab:elHam}
\begin{tabular}{ccccccccc}
    & $ ^1(S_0 S_0)       $
    & $ ^1(T_1 T_1)       $
    & $ ^1(S_0 S_1)       $
    & $ ^1(S_1 S_0)       $
    & $ ^1(CA)            $
    & $ ^1(A C)           $
    & $ ^1(D E)_1         $
    & $ ^1(D E)_2         $ \\
 \hline
 \hline
    $ ^1(S_0 S_0)$ & -174.3 &   23.0 &  -28.5   &   36.6  &   19.2   &  -16.2 &   377.9 &   380.3 \\
    $ ^1(T_1 T_1)$ &   23.0 & 1951.8 &    4.4   &   -8.8  & -202.5   &  200.5 &   -16.3 &   -13.3 \\
    $ ^1(S_1 S_0)$ &  -28.5 &    4.4 & 1564.4   &  -36.3  & 13.9     &   24.2 &    39.9 &     9.0 \\
    $ ^1(S_0 S_1)$ &   36.6 &   -8.8 &  -36.3   & 1555.7  & 26.4     &   14.0 &   -12.6 &   -43.8 \\
    $ ^1(C A)$     &   19.2 & -202.5 &   13.9   &   26.4  & 1827.7   &   27.2 &    54.9 &    37.3 \\
    $ ^1(A C)$     &  -16.2 &  200.5 &   24.2   &   14.0  & 27.2     & 1829.6 &   -38.3 &   -53.4 \\
    $ ^1(D E)_1$   &  377.9 &  -16.3 &   39.9   &  -12.6  & 54.9     &  -38.3 &  2135.2 &  -126.2 \\
    $ ^1(D E)_2$   &  380.3 &  -13.3 &    9.0   &  -43.8  & 37.3     &  -53.4 &  -126.2 &  2113.5
\end{tabular}
\end{table}

\subsection{MCTDH Dynamics and Analysis}

The time-dependent Schrödinger equation for the molecules-cavity system
was integrated using the MCTDH approach~\cite{man_92_3199, beck2000multiconfiguration}
in its multilayer
generalization~\cite{wan_03_1289,man_08_164116,vendrell_multilayer_2011}
using the MCTDH Heidelberg package~\cite{mctdh:MLpackage}.
The treatment of molecular ensembles coupled to cavity modes using the ML-MCTDH
approach has been described in some detail in
Ref.~\citenum{vendrell_coherent_2018} and used in several
applications~\cite{vendrell_coherent_2018,ulusoy_modifying_2019}.
In particular, it is most convenient when considering
an ensemble of molecules that each molecule has its own separate electronic and
vibrational degrees of freedom. This allows for layering schemes of the ML tree
where each molecule "lives" in a separate sub-tree, the advantage being that the
strongest correlations are within the molecules themselves compared to the
cavity-molecules coupling.
Examples of the tensor trees used in the simulations are found in the
Appendix. The primitive degrees of freedom of each molecule are treated using
the same grids as in Ref.~\citenum{reddy_intramolecular_2018}.

In order to follow the dynamics of the coupled electronic and cavity states we
define the upper and lower polaritons using their amplitudes at exact resonance
\begin{align}
    |\psi_{\pm}\rangle=
    \frac{1}{\sqrt{2}} |1, 0\rangle
    \pm
    \frac{1}{\sqrt{2N}}
    \sum_{k=1}^{N}
        |0, k_+\rangle.
\end{align}
Here $|1, 0\rangle$ indicates an excitation in the cavity while no molecules are
excited and $|0, k_+\rangle$ corresponds to the unexcited cavity and a bright
exciton in the $k$-th molecule.
Specifically, since each molecule corresponds to an excitonic dimer, one can
introduce the bright and dark local excitations
\begin{align}
    |0, k_\pm\rangle = \frac{1}{\sqrt{2}}
    \left(
        |0, k_{10}\rangle \pm |0, k_{01}\rangle
    \right)
    \label{eq:excitons}
\end{align}
where $|0, k_{10}\rangle$ corresponds to the $k$-th dimer with the first
monomer excited, and so on.
Following these definitions one can introduce projectors onto the
specific subspaces, for example
$\hat{P}_{\pm}=|\psi_{\pm}\rangle\langle\psi_{\pm}|$.

When simulating the dynamics triggered by an explicit laser pulse we define
triplet-triplet yield of dimer $k$ as
\begin{align}
    \mathrm{Y}_k=\frac{P_\text{ME}^{(k)}(t)}{1-P_{00}^{k}(t)},
    \label{eq:SF_yield_k}
\end{align}
where $P_\text{ME}^{(k)}(t)$ indicates the population of the
final ME state of dimer $k$ at time $t$ and $P_{00}^{k}(t)$ is the
corresponding population of the electronic ground state.
Since all dimers are identical, and the dynamics takes place within
the single-excitation manifold, it follows that
\mbox{$\text{Y} = N\cdot\text{Y}_k$}.

Alternatively, one can start the dynamics either initially populating the
cavity, populating the molecules coherently, or populating either the upper or
lower polaritons. For the latter, one applies the transition
operator~\cite{ulusoy_modifying_2019}
\begin{align}
    \hat{T}_{\pm} =
    \frac{1}{\sqrt{2}}
    \left(\hat{a}^{\dagger}
    \pm
    \frac{1}{\sqrt{N}}
    \sum_{k=1}^N
    \left(
    \frac{\hat{\sigma}^{\dagger}_{k,01}
        + \hat{\sigma}^{\dagger}_{k,10}
        }
        {\sqrt{2}}\right)\right)
    \label{eq:Pol_operators}
\end{align}
where the ladder operators above are defined by
\begin{align}
    \hat{a}^{\dagger} |0,0\rangle & = |1, 0\rangle \\
    \hat{\sigma}^{\dagger}_{k,01} |0,0\rangle & = |0,k_{01}\rangle \\
    \hat{\sigma}^{\dagger}_{k,01} |0,0\rangle & = |0,k_{01}\rangle.
    \label{eq:ladder_operators}
\end{align}
Absorption spectra were calculated by Fourier transform of the
autocorrelation function of the dipole-operated ground state or the
ground state operated with some specific transition operator as defined above.

\section{Results and Discussion}

\subsection{Modified Intramolecular Singlet Fission in the \textit{o}-TIPSm dimer}

\begin{figure}[t]
   \includegraphics[width=8cm]{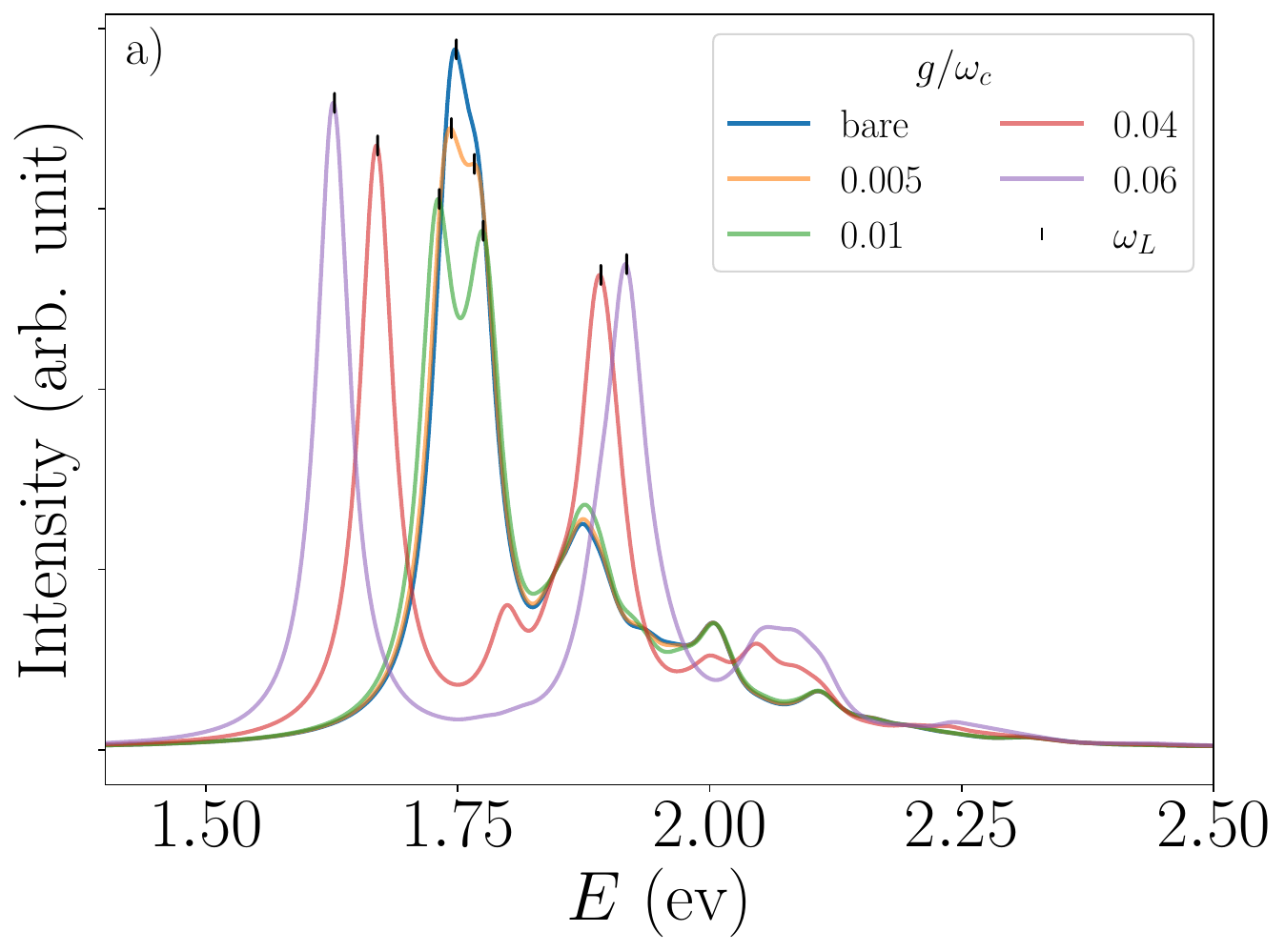}
   \includegraphics[width=8cm]{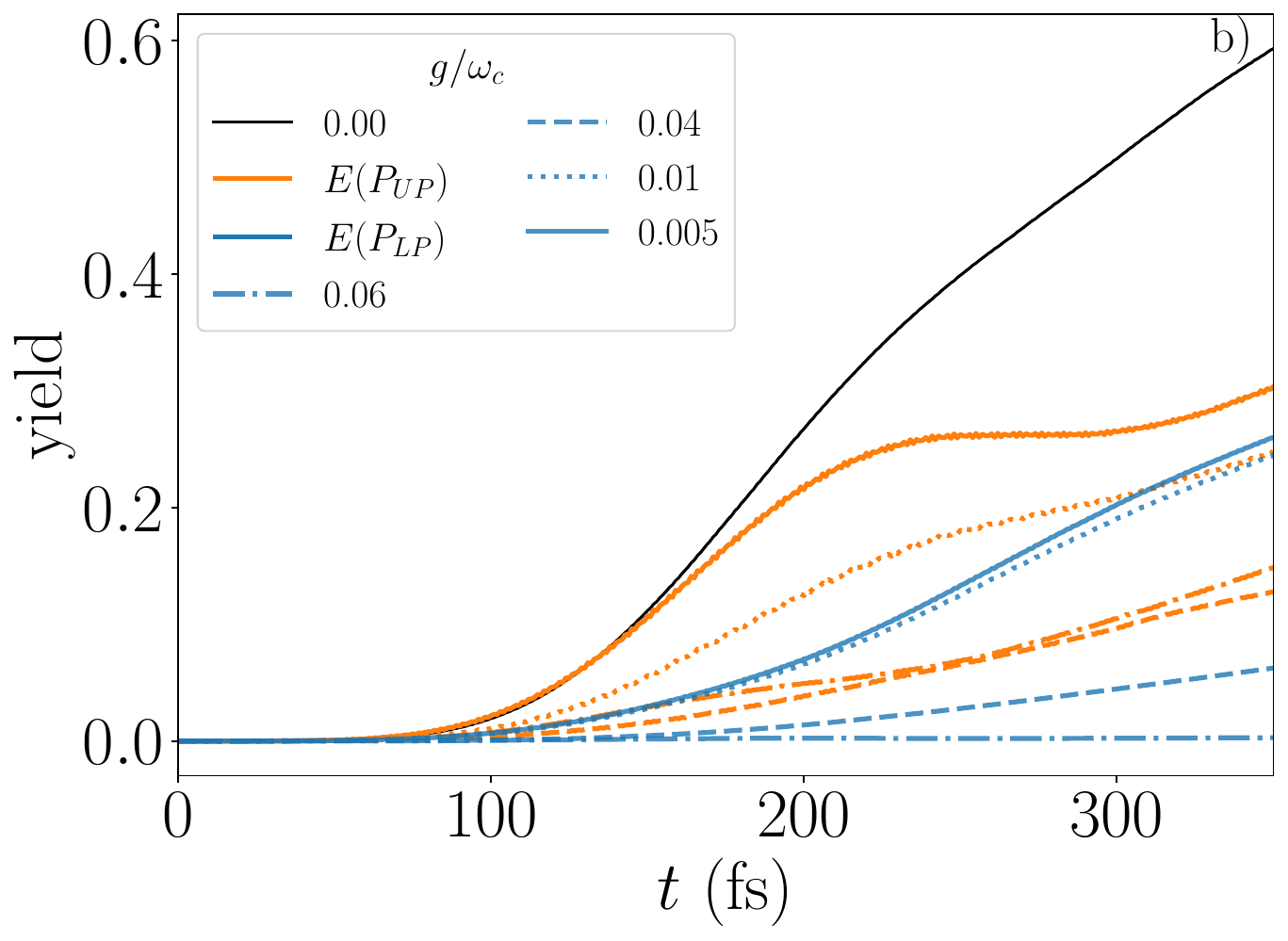}
    \caption{(a) Absorption spectra of the bare \textit{o}-TIPSm
    dimer and the dimer coupled to the cavity at various coupling strengths.
    The small vertical lines indicate the photon energy ,$\omega_L=$(1.626~eV, 1.669~eV,1.730~eV, 1,743~eV, 1,747~eV, 1.765~eV, 1.774~eV, 1.891~eV, 1.915~eV) from left to right, of the
    corresponding simulations targeting the bright absorption lines.
    (b) Comparison of the singlet fission yield of the bare molecule
    and the cavity-coupled molecule for different coupling strengths after laser
    excitation. The blue lines represent the yield after excitation to the UP
    state, while the orange lines represent the yield after excitation to the LP
    state. The fast Rabi oscillations were smoothed by taking a running average
    over 10 femtoseconds.}
    \label{fig:yield_good_sys}
\end{figure}

First we consider a single dimer coupled to the cavity and four
coupling strengths $\tilde{g}/\omega_c = (5\cdot 10^{-3}, 1\cdot 10^{-2}, 4\cdot 10^{-2}, 6\cdot 10^{-2})$.
We note that the $\tilde{g}$ is assumed to be already multiplied with the
molecular transition dipole matrix element, and in the following we refer to
this parameter as $g$.
For each coupling strength, the corresponding electronic
absorption spectrum calculated from the autocorelation function of the
dipole-operated ground states in shown in Fig.~\ref{fig:yield_good_sys}a.
It can be seen how the Rabi splitting increases with the coupling strength,
resulting in clearly separated UP and LP resonances for
${g}/\omega_c = 4\cdot 10^{-2}$ and larger couplings.
These absorption lines, indicated by the small vertical lines in
Fig.~\ref{fig:yield_good_sys}a, were targeted by a laser pulse with
carrier frequency resonant with the center of the corresponding absorption
band and spectral bandwidth (fwhm) of 0.02~eV and the duration of the laser pulse is set to 90~fs. 
The amplitude of the laser was set to ensure that the excitation remained within
the one-photon regime~\cite{ulusoy_modifying_2019}.

The different propagations initiated by the explicit laser pulses targeted the
UP and LP bands at all coupling strengths. The cavity was set in all cases
resonant with the first singlet excitation of the dimer at $\omega_c=1.75$~eV
and the laser frequencies are given in the caption of Fig.~\ref{fig:yield_good_sys}.
Irrespective of the coupling strength of the cavity and the laser frequency
$\omega_L$, coupling to the cavity decreased the population transfer to the
final triplet-triplet ME state in all instances compared to the bare system, as
seen in Fig.~\ref{fig:yield_good_sys}b.

 \begin{figure}[t]
    \centering
    \includegraphics[width = 0.5\textwidth]{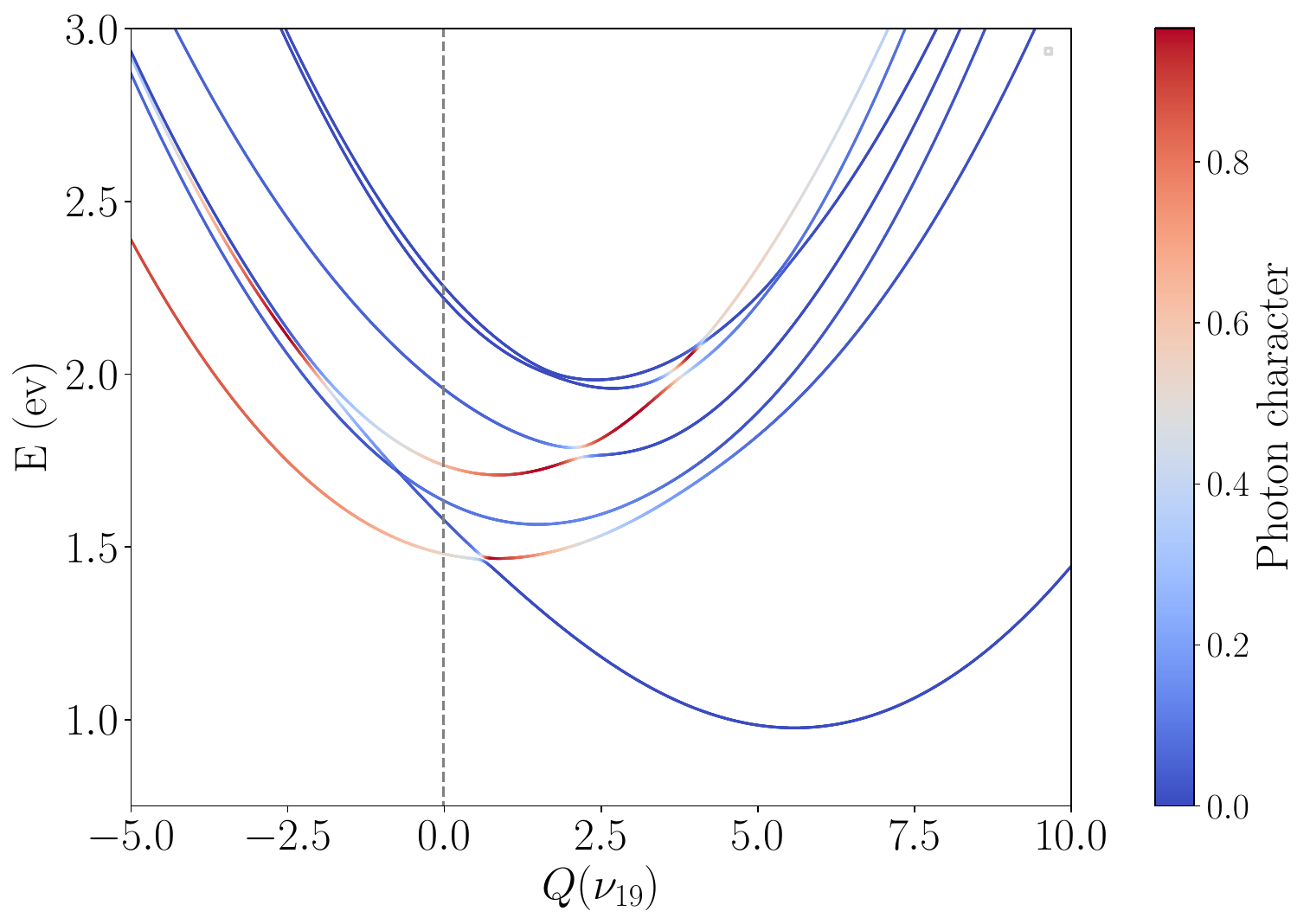}
    \caption{Cut of the adiabatic polaritonic surface along mode $Q_{19}$ with a coupling strength of 0.06 where the gradient indicates the character of the state. The gray dashed
     line shows the position of the Franck-Condon point. For the sake of
     clarity, the ground state is not shown in this plot. }
    \label{fig:adiabatic_cut}
 \end{figure}

The effect of the cavity on the electronic structure of the dimer
turns out to be not so simple. Inspecting the cut of the adiabatic PESs along
the tuning vibrational mode $Q_{19}$, Fig.~\ref{fig:adiabatic_cut},
we can identify two parallel polaritonic
PES with substantial photonic character, the LP and UP, which cross the curve
leading to the minimum energy geometry of the ME state, but also present avoided
crossings with the intermediate adiabatic states.
In order to better understand the dynamics leading to the reduced yield, we
start simulations selectively from the UP and LP states using the transition
operators in Eq.~\ref{eq:Pol_operators}, and compare them to the time evolution
of the bare system started from the bright excitonic state of the dimer.
In the analysis, the electronic-photonic population is split into three
subspaces. The first subspace contains only the final diabatic ME state. The
second subspace contains all diabatic states directly involved with the
light-matter interaction, namely the ground and singlet excited electronic
states of the dimer plus the one-photon excitation of the cavity (in the bare
system the latter state is obviously absent). The third subgroup encompasses the
four intermediate DE and CT electronic states.

\begin{figure}[t]
    \centering
        \includegraphics[width=\textwidth]{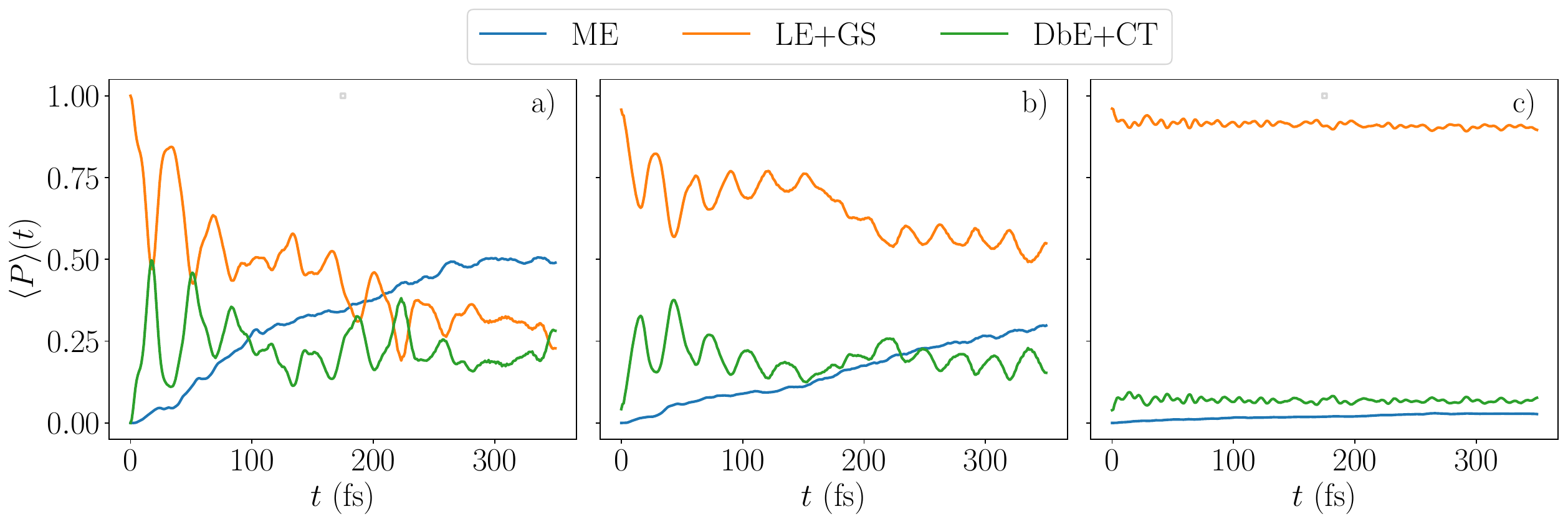}
    \caption{Time evolution of the electronic populations within
    the dimer. (a) bare and (b, c) cavity-coupled system
    at coupling strength 0.06.
    In each case the initial state corresponds to
    (a) the bright exciton $|0, 1_+\rangle$,
    (b) the UP $|\psi_+\rangle$ and
    (c) the LP $|\psi_-\rangle$.}
    \label{fig:popelectronicstates}
\end{figure}

In the bare system, photo-excitation of the singlet resonance results in quick
population transfer to the intermediate states (green trace in
Fig.~\ref{fig:popelectronicstates}a) which subsequently drifts towards the final
ME state.
Exciting the UP results in a reduced population transfer to the intermediate
states, and between 50 to 70\% of the population stays trapped in the
polaritonic subsystem (orange trace in Fig.~\ref{fig:popelectronicstates}b), and
would ultimately decay as photon losses~\cite{ulu_20_044108}.
Excitation of the LP state results in a complete suppression of the population
transfer to the intermediate states, which are then too high in energy with
respect to the lower polaritonic excitation.
This example illustrates the energy shift effect of the
polaritonic excitations due to Rabi splitting when compared to the bare
system. The effect of the cavity remains relevant even when the time-scale
for reaching the final ME state is long compared to typical cavity lifetimes of
tens of femtoseconds.
The reason for this is that the cavity resonance merely produces the doorway state
through which the external light enters the system, and the passage through this state towards the various intermediate states
may be shorter than the cavity lifetime.

\begin{figure}[t]
    \centering
    \includegraphics[width=1\textwidth]{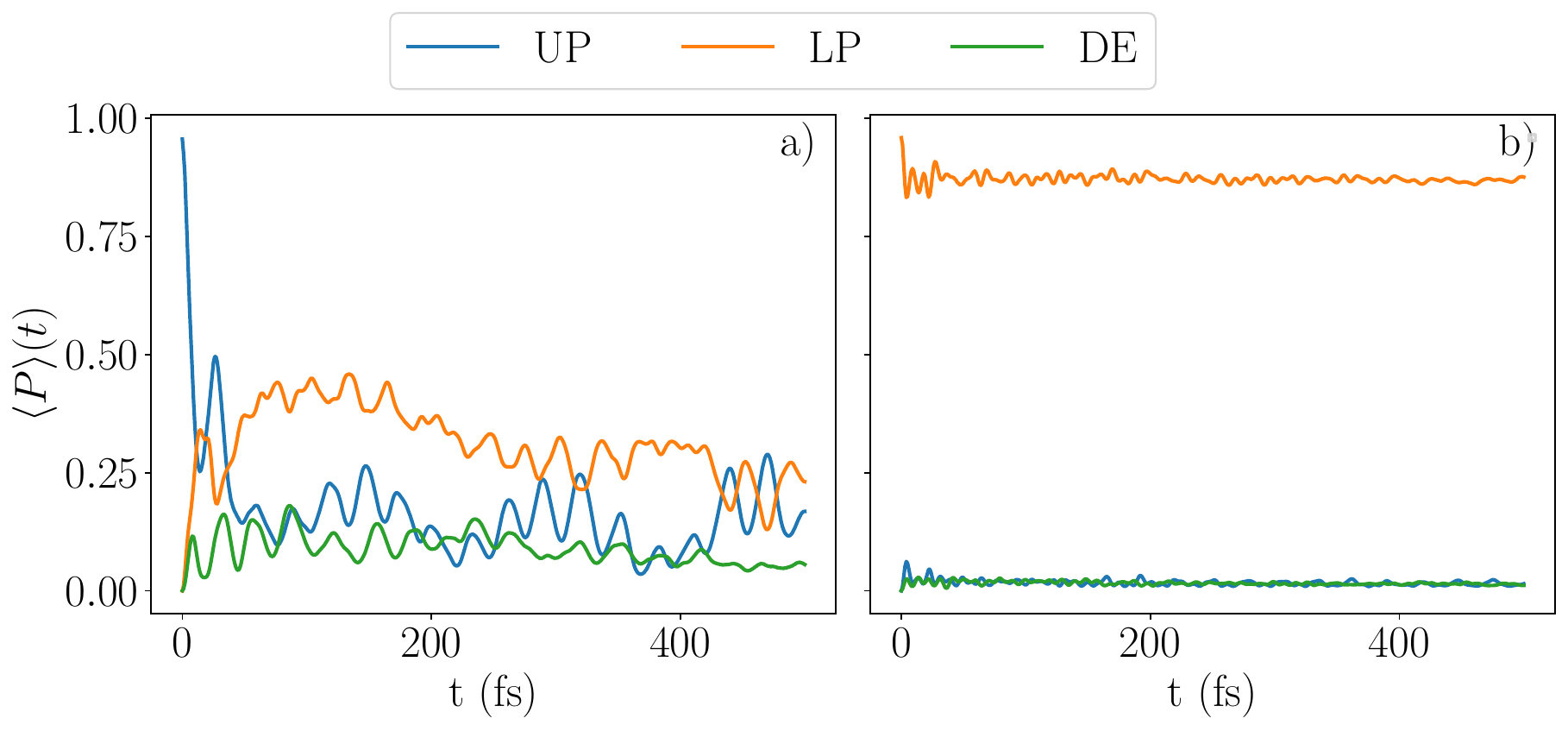}
    \caption{Population of the UP, LP and dark exciton states as a function
             of time following the excitation to (a) the UP and (b) the LP
             states of a single dimer coupled to the cavity. The simulations are
             the same as in Fig.~\ref{fig:popelectronicstates}.}
    \label{fig:pop_polaritons}
\end{figure}

A closer look at the dynamics starting from the UP and LP states is provided in
Fig.~\ref{fig:pop_polaritons}.
Following the excitation of the UP, most of the
population rapidly transitions to the LP and the dark exciton
(cf. Eq.~\ref{eq:excitons}), where it becomes trapped.
Following the excitation of the LP, and due to its too low energy in relation to
the intermediate states, the population remains trapped and the SF process is
then completely quenched.

\subsubsection{Collective Effects}

\begin{figure}[t]
    \centering
    \includegraphics[width = \textwidth]{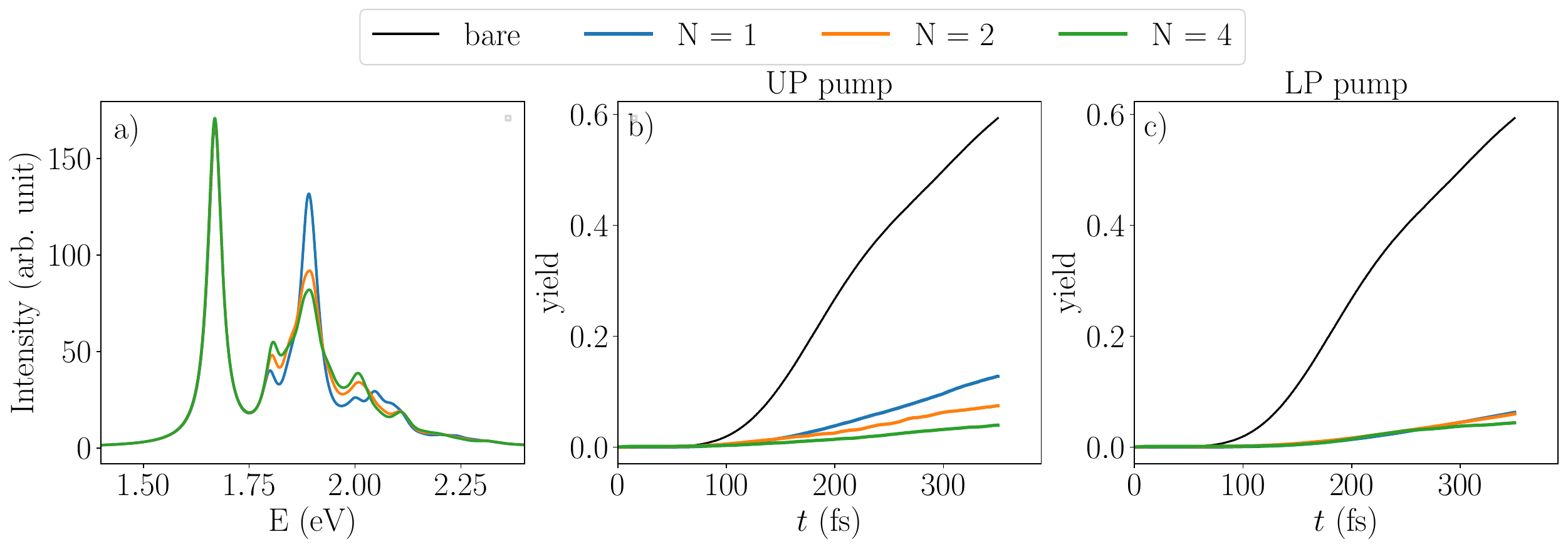}
    \caption{(a) Absorption spectra for a set light-matter coupling strength for
    one, two and four dimer molecules in the cavity with $g/{\omega}=0.04$.
    Calculated TT yield after pumping the UP (b) or the LP (c)  for 1,2 and 4
    molecules. The fast Rabi oscillations where smoothed by a running average
    over 10 data points}
    \label{fig:multimols_good}
\end{figure}

 Finally, we consider an ensemble of $N$
 molecules (in this case $N$ dimers) and as usual scale down the coupling to the
 cavity by the factor $N^{-1/2}$ to maintain a constant Rabi splitting, and thus
 as constant energy of the bright doorway states, in all simulations.
 The \textit{o}TIPS dimer model was extended to to include $N=2$ and
 $4$ molecules.
 The absorption spectrum for the cases $N=1,2,4$
 (cf. Fig. \ref{fig:multimols_good}a) is virtually identical in the
 region of the LP resonance, whereas the UP resonance becomes broader as the
 number of coupled molecules is increased. This broadening is caused by the well-known
 involvement of the nominally dark polaritonic states, non-adiabatically
 coupled to the bright states~\cite{vendrell_collective_2018}.
 A broadening effect in the UP is equally seen in vibrational strong
 coupling~\cite{gomez_vibrational_2023},
 and can be understood as being generated by resonances involving the LP and
 dark states plus vibrational excitations and the UP.
 These couplings
 result in a fast non-radiative relaxation from the UP towards the dark
 states and the LP.

 Correspondingly, the SF yield, i.e. the final ME state population,
 when starting from the LP is quite insensitive to collective effects. The LP is
 in all cases off-resonant with the intermediate states and the population
 remains trapped there, as seen in Fig. \ref{fig:multimols_good}c.
 The dynamics starting from the UP are instead sensitive to $N$ because the
 decay rate towards the dark states and LP depends on the number of
 molecules~\cite{vendrell_collective_2018}. In this case,
 the system relaxes towards the LP and dark states faster for
 larger $N$.
 Interestingly, the fast relaxation towards the dark states
 manifold, which are energetically equal to the bare dimer excitation,
 does not result in an increased ME yield.
 This can be understood from the perspective that
 the initial excitation energy of the UP quickly redistributes
 among the vibrational modes as suggested by the broader absorption resonance,
 thus leading after a few tens of femtoseconds to a similar energetic
 situation as reached through the direct excitation of the LP.

\subsection{Upper Polariton as a Doorway State}

 These results seem to imply that using the UP as a doorway state for SF cannot
 improve the SF yield.
 However, the situation in which the LE excitations of the bare system are too
 low compared to the $^1(T_1T_1)$ state is rather
 common~\cite{padula_singlet_2019}, and in this case a modified resonance
 condition that uses the UP to couple to the final ME might enhance the
 yield.

\begin{figure}[t]
 \centering
    \includegraphics[width=0.8\textwidth]{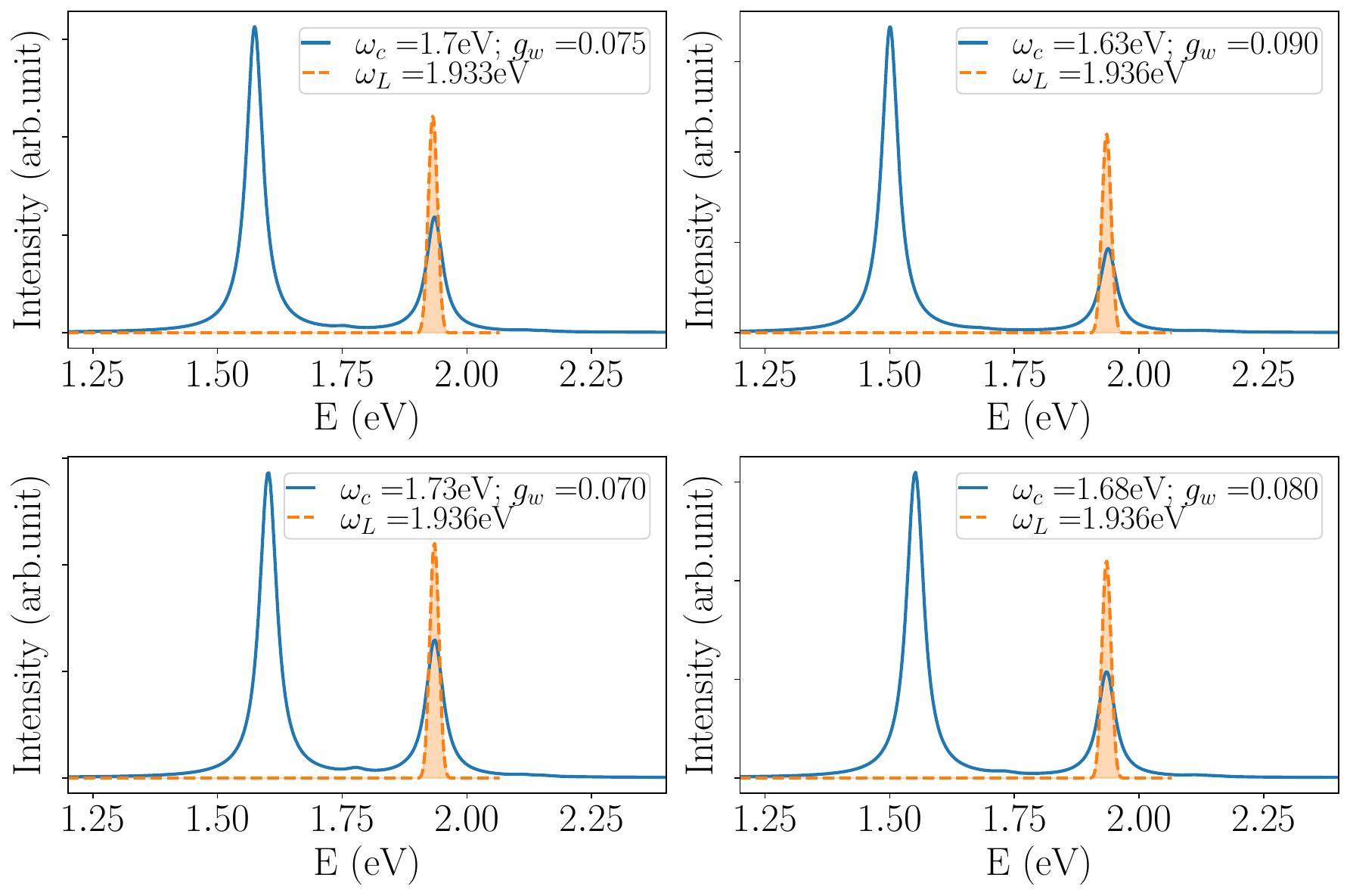}
    \caption{Absorption spectra for some parameter combinations, each with the
    corresponding laser directed at the UP.}
     \label{fig:scan_para}
 \end{figure}

 In order to investigate this possibility thoroughly, we reduced the
 oTIPSm to only the main tuning $Q_{19}$ mode but kept all
 8 electronic states of each dimer (This model corresponds to the
 simulations shown in Fig. 5b in Ref.~\citenum{reddy_intramolecular_2018}).
 We then shifted the energy of the LE states down by about 0.1~eV until no noticeable
 population transfer to the ME state occurs in the bare system.
This corresponds to an energy gap of 1.73~eV between the
 electronic ground state and the LE excitation.
 For this model,
 the cavity parameters were varied within the frequencies
 $1.6-\SI{2.0}{\electronvolt}$ in \SI{0.01}{\electronvolt} steps
 and the light-matter coupling strength was varied in the range $0.045-0.090$
 in $0.005$ steps, thus resulting in 396 parameter combinations.
 Each parameter combination underwent relaxation to the ground
 vibro-polaritonic
 state, and a corresponding absorption
 spectrum was calculated to determine the exact position of the UP resonance.
 The frequency of the laser $\omega_L$ used to resonantly couple to
 the UP state was then set in each case to correspond to the
 excitation energy of the center of the UP absorption band.
 The absorption spectrum corresponding to a few selected parameter combinations
 together with the spectral distribution of the excitation pulse are shown in
 Fig.~\ref{fig:scan_para}.

\begin{figure}[H]
    \centering
    \includegraphics[width=0.9\textwidth]{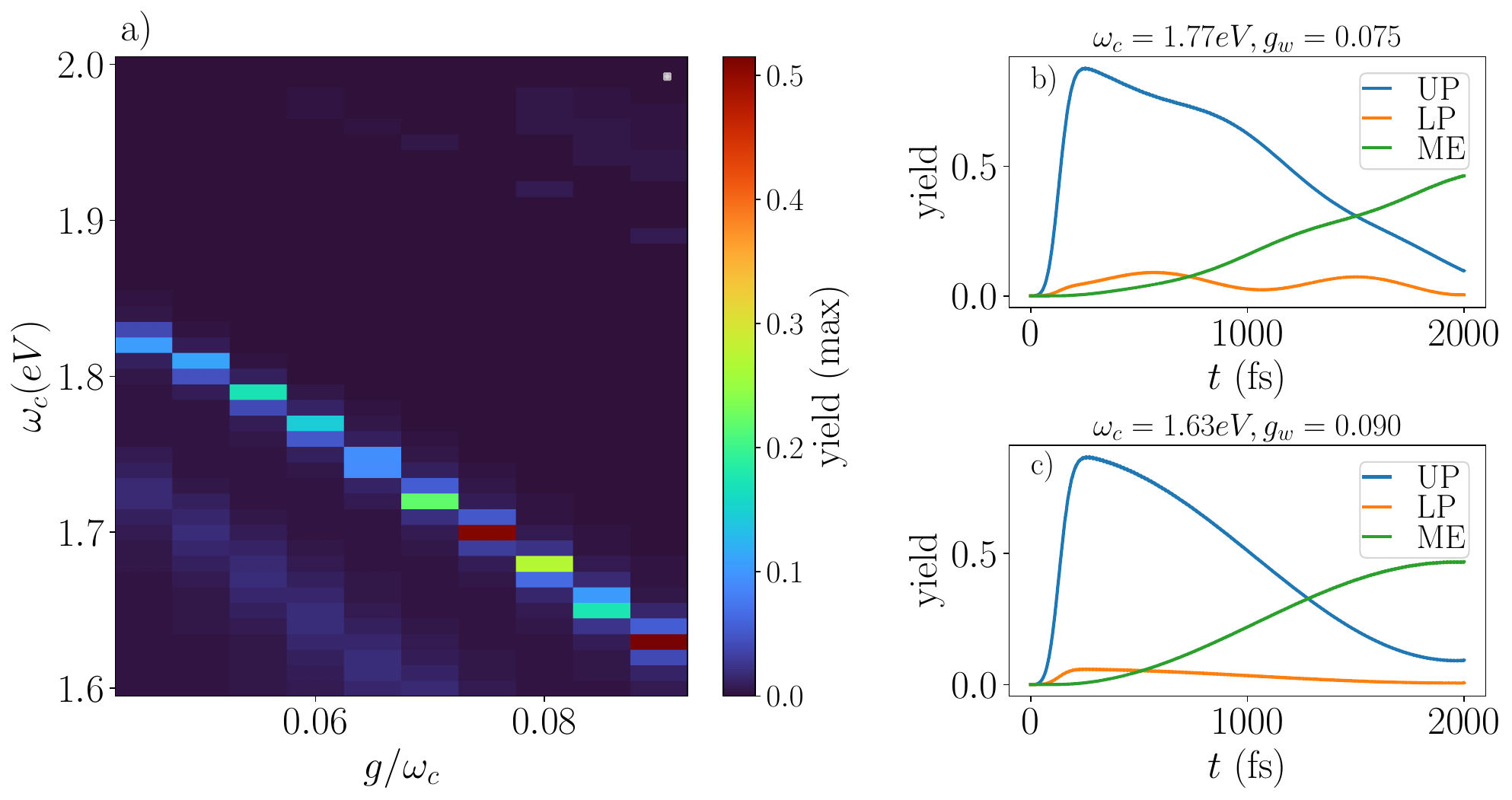}
   \caption{Calculations for the non-optimal singlet fission system for one
    molecule, (a) shows a systematic scan of different cavity frequencies and
    coupling strengths and the corresponding ME yield when pumping the UP.(b, c) Time-dependent energy transfer dynamics between the UP, LP
    and ME states for specific parameters. The fast Rabi oscillations where smoothed by a running average over 10 data points.}
    \label{fig:scan1M}
\end{figure}

 Most of the parameter space results in close to zero yield, as seen in
 Fig.~\ref{fig:scan1M}a.
 Let us first consider the cavity frequency $\omega_c=1.7$~eV, the closest
 value in the parameter space to the resonance frequency of the singlet
 resonance. Only when the coupling strength reaches
 about $g/\omega_c=0.075$ does the population of the ME state reaches a
 significant percentage of about 50\%. This corresponds to the UP being at the
 ideal energy to act as a doorway toward the ME final state.
 If the cavity is tuned toward higher frequency, a correspondingly smaller
 coupling strength is needed to reach the same optimal UP energy, as
 illustrated by the nearly linear trend in the yield in Fig.~\ref{fig:scan1M}a.
 Since the cavity energy is above the singlet excitation energy, the UP is more
 photonic than molecular, and the total yield reaches only about 15 to 20\%.
 On the other hand, when the cavity is tuned below the LE resonance in the
 region of 1.6 to 1.7~eV, an increasingly larger coupling strength is required
 to bring the UP to its optimal energy for the SF mechanism to operate.
 However, when this
 condition is achieved, the UP has now a larger molecular than photonic
 character, which leads to a larger yield compared to the cavity with a higher
 frequency.
 Figures~\ref{fig:scan1M}b and \ref{fig:scan1M}c show the time-dependent
 population of the UP, LP and ME states for the parameter combinations with the
 largest yield.

\subsection{Collective Effects in the UP Doorway Mechanism}

\begin{figure}[t]
    \centering
    \includegraphics[width=\textwidth]{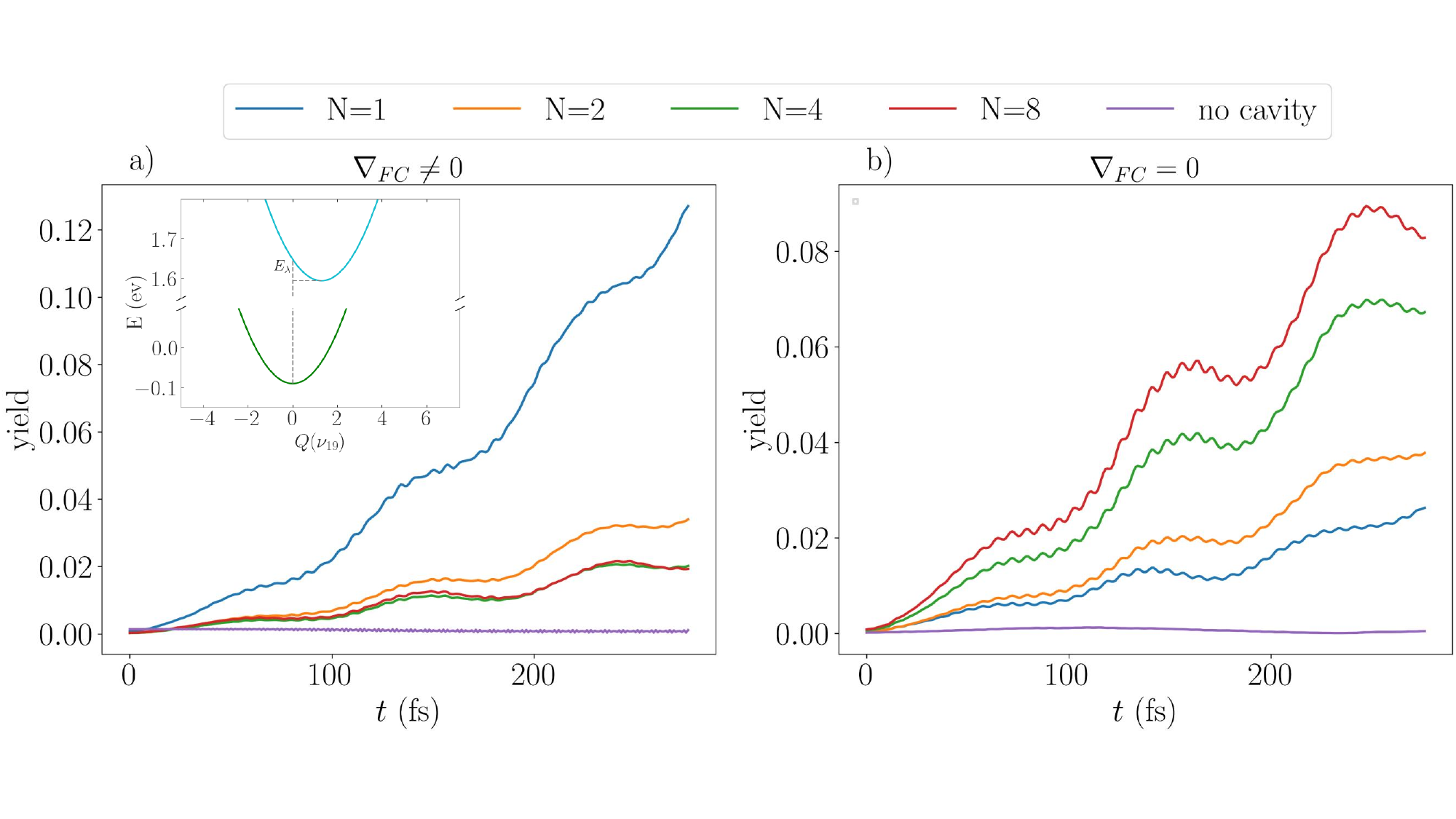}
    \caption{Time evolution of the $^1(T_1T_1)$ yield after an instantaneous
    photoexcitation of the UP for different numbers of molecules ($N$) for the
    two non-optimal iSF model systems with (a) gradient at the FC point and (b)
    no gradient at the FC point. The diabatic representation of the first
    excited state along the vibrational mode and the reorganizations energy are
    shown on the upper right. The fast Rabi oscillations where smoothed by a
    running average over 25 data points. }
    \label{fig:multipMols}
\end{figure}

As mentioned above and as it has been thoroughly
analyzed~\cite{vendrell_collective_2018}, the UP states has an extra open decay
channel toward the manifold of $N-1$ DS and the LP in the presence of
$N$ molecules. The influence of this channel on the UP doorway mechanism and
possible ways to overcome its negative effect need to be analyzed.
When considering this scenario, we weight the coupling strength as usual
with $N^{-1/2}$ so that the Rabi splitting is kept constant as a function
of $N$.
Figure~\ref{fig:multipMols}a illustrates the population of the final ME state
for the optimal cavity coefficients determined for one molecule and
for an increasing number of molecules, where a converged population
dynamics has been reached with $N=8$. The decay channel towards the
DS and LP very clearly quenches the SF yield.
The efficiency of this relaxation pathway is proportional to the gradient
difference between the two electronic states coupled via the cavity, as
illustrated in the inset of Fig.~\ref{fig:multipMols}a.
Once the two electronic states
become coupled by the cavity, the displaced position of the minima on both
PES results in the resonance condition becoming dependent on the displacement of
the vibrational modes, which results in the vibro-polaritonic coupling.

A natural way to counter this decay channel is to demand that the PESs of the
electronic ground state and the LE state coupled by the cavity are
as parallel as possible.
This is a problem of molecular structure and design that has to be solved by
considering adequate modifications of the specific molecular scaffolds
in chemical space~\cite{dre_23_17079}.

Here we consider the modification of the model by shifting the ground
electronic state horizontally along the vibrational mode until the gradient
difference vanishes.
This case is illustrated in Fig.~\ref{fig:multipMols}b. First of all we see here
that the $N=1$ case has a substantially smaller yield that in the non shifted
case. We should note that all electronic states are coupled to each other
(cf. Table~\ref{tab:elHam}) and that shifting the ground state fundamentally
changes the model, such that direct comparisons are no longer possible.
Now, however, the decay pathway from the UP towards the DS and LP is not
operative and the yield does not decrease with $N$. Instead, one observes
that, in this model, the yield increases with $N$. One must take this with a
grain of salt because as just mentioned, all states including the intermediate
CT and DE states are coupled among each other. As the ground state has been
moved parallel to the LE states, other side effects may have occurred leading to
the observed behaviour.
What one can clearly conclude is that a small gradient difference between the
cavity coupled states permits the UP state to remain operative as the doorway
state, and that realizing this mechanism becomes
a question of finding adequate molecular systems.

\section{Conclusions}

We considered process of intramolecular singlet fission
in \textit{o}-TIPSm dimer, previously studied theoretically by
Reddy and Thoss~\cite{reddy_intramolecular_2018}, under the effect of
strong light-matter coupling in a cavity.
This system presents a complex SF mechanism involving the participation of
several intermediate CT and DE states mediating the population transfer
the final ME state.
Simulations in the cavity
indicate that the modification of the excitation
energy of the doorway state, the LP and UP in the case of strong coupling,
result in a large suppression of the ME yield. This is particularly dramatic for
the LP, which at moderate coupling strengths is found too low in energy to
result in any significant population transfer toward the intermediate states and
finally toward the ME state.
The results indicate that the energy of the optically bright states, those
directly affected by the cavity, relative to the intermediate states, which are
insensitive to it, is the relevant quantity that determines the effectivity of
the mechanism.

Subsequently we considered the situation where the LE state is too low compared
to the $^1(T_1T_1)$ excitation. This is a common situation that invalidates a
large set of molecular structures as SF candidates, but which could be remedied
by strong coupling and using the UP as doorway state.
An exhaustive cavity parameter analysis indicates that it is possible to find
cavity frequency and coupling strength combinations that can largely restore the
SF yield in this case. However, using the UP as the doorway state has the caveat
that the fast decay towards the dark states and the LP can jeopardize this
possibility.
We thus consider the situation in which the non-radiative decay
channel from the UP is suppressed by reducing the gradient difference along the
vibrational modes between the ground electronic state and the LE state.
In this case, the SF mechanism can still operate using the upper
polaritonic resonance.
It then becomes a search problem in chemical space to identify molecular scaffolds
fulfilling, at least to some extent, this condition.

\begin{acknowledgement}

O.V. acknowledges the collaborative research center
"SFB 1249: N-Heteropolyzyklen als Funktionsmaterialen" of
the German Research Foundation (DFG) for financial support.

\end{acknowledgement}

\bibliography{SingletFission}

\newpage
\appendix
\section{Appendix}

\section{Calculation details}

\begin{figure}[H]
    \centering
    \includegraphics[width=0.5\textwidth]{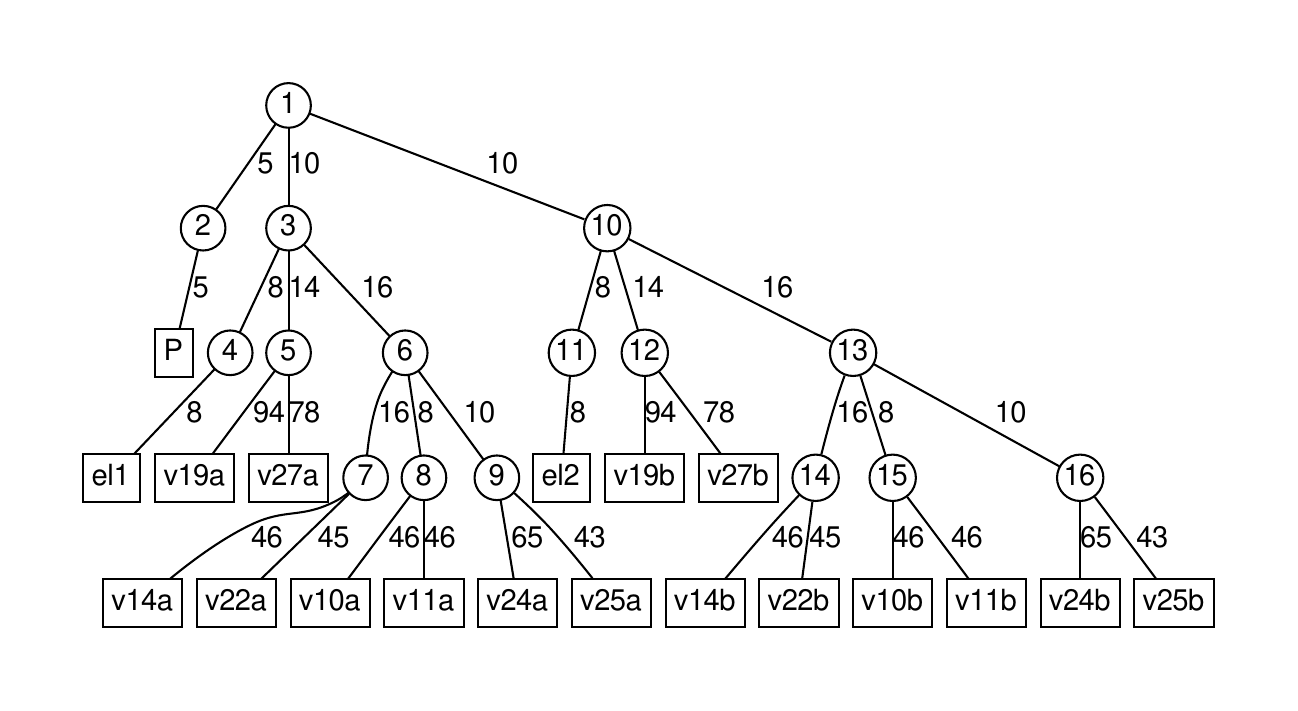}
     \includegraphics[width=0.8\textwidth]{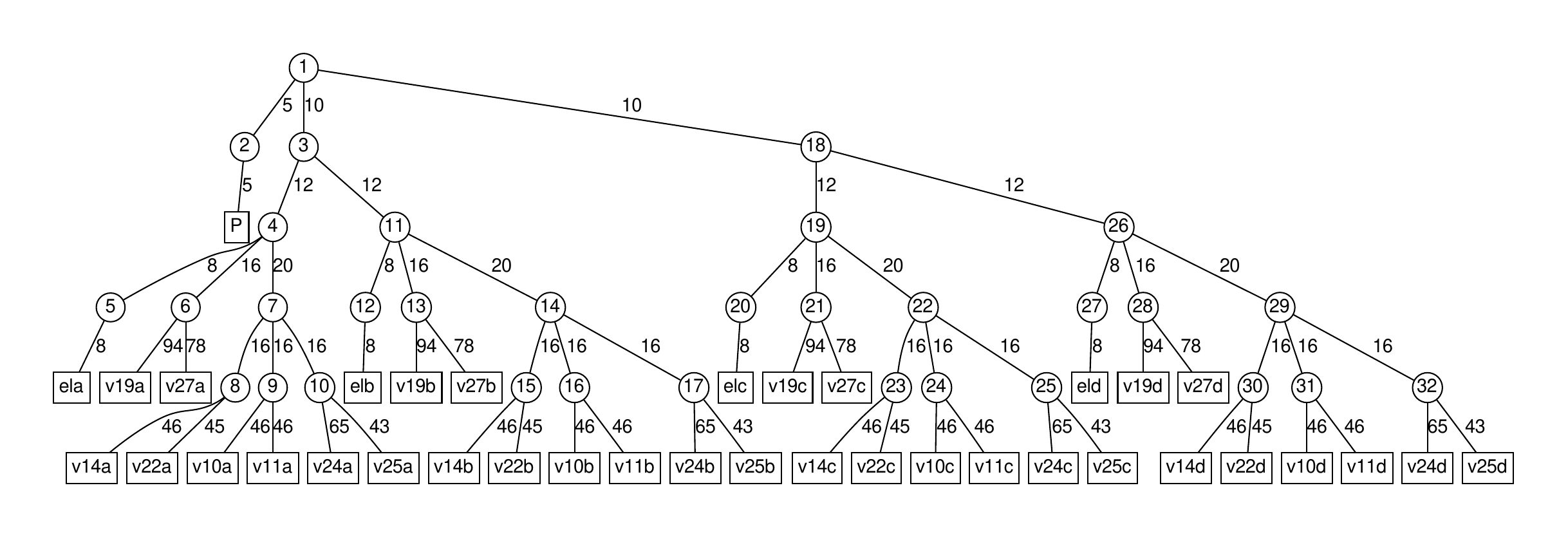}
    \caption{ML-tree structure of the \textit{o}-TIPS model for two (a) and four (b) molecules. The number of each node can be seen in the circles. The number of SPFs per node are given by the number along the branches. The square boxes indicate which degree of freedom it represents and above them the number of primitive basis functions is given. }
    \label{fig:mltrees_good}
\end{figure}

Figure \ref{fig:mltrees_good} shows the ML-tree structures for
the \textit{o}-TIPS model for two and four molecules.
Each molecule has separate vibrational and electronic degrees of freedom.
 All vibrations are represented by a harmonic oscillator DVR using the same
 grids as in Ref.~\cite{reddy_intramolecular_2018}.
 The electronic degrees of freedom and the cavity are represented by
 general discrete bases.

\end{document}